\documentclass[prb,showpacs,superscriptaddress,twocolumn]{revtex4}%
\usepackage{amsfonts}
\usepackage{amsmath}
\usepackage{amssymb}
\usepackage{graphicx}%
\begin{document}
\title{Magnetization in two-dimensional electron gas in a perpendicular magnetic
field: the roles of edge states and spin-orbit coupling}
\author{Zhigang Wang}
\affiliation{LCP, Institute of Applied Physics and Computational Mathematics, P.O. Box
8009, Beijing 100088, People's Republic of China}
\author{Wei Zhang}
\affiliation{LCP, Institute of Applied Physics and Computational Mathematics, P.O. Box
8009, Beijing 100088, People's Republic of China}
\author{Ping Zhang}
\thanks{Author to whom correspondence should be addressed. Email
address: zhang\_ping@iapcm.ac.cn} \affiliation{LCP, Institute of
Applied Physics and Computational Mathematics, P.O. Box 8009,
Beijing 100088, People's Republic of China} \affiliation{Center for
Applied Physics and Technology, Peking University, Beijing 100871,
People's Republic of China} 

\begin{abstract}
We study the de Haas--van Alphen (dHvA) oscillations in the magnetization of a
two-dimensional electron gas (2DEG) under the influence of the edge states
and/or the Rashba spin-orbit interaction (SOI). The boundaries of the systems
lift partially the degeneracies of Landau levels (LL's) and the resulting edge
states lead to the changes of both the center and the amplitude of the
sawtoothlike magnetization oscillation. The SOI mixes the spin-up and
spin-down states of neighboring LL's into two unequally spaced energy
branches.
The inclusion of SOI changes the well-defined sawtooth pattern of the dHvA
oscillations in the magnetization.
The weaker the magnetic field is, the larger is the change of the dHvA
oscillations due to the edge effect and/or the spin-orbit coupling. Some
theoretical results are compared with the experimental data.

\end{abstract}
\maketitle

\section{Introduction}

The physics of two-dimensional electron gas (2DEG) in the presence of a
perpendicular magnetic field reveals a rich variety of remarkable phenomena,
for example the integer and fractional quantum Hall effects (QHEs). Recently,
the magnetization property of 2DEG at low temperature and in a strong
perpendicular magnetic field has attracted extensive interest.
It is due to the fact that the magnetization is particularly suited to
investigate the electronic ground-state properties and the density of states
(DOS) of 2DEG by its minimal perturbation to the system \cite{book}. On the
experimental side, the measurement of the magnetization keeps a most
challenging task due to weak signal of the magnetization. A variety of
techniques, such as dc superconducting quantum interference device (SQUID)
magnetometers \cite{Stormer1983,Mei1997,Mei1999,Mei2003}, picking up coils
lithographed above the gate \cite{Fang1983}, torque magnetometers
\cite{Eisen1,Eisen2,Wieg1997,Zhu2003,Matthews2004,Faul2005}, torsional
magnetometer with optical angular detection
\cite{Schaapman2002,Schaapman2003,Bom2006}, and micromechanical cantilever
magnetometers \cite{Sch1,Sch2,Wilde2006,Kno2002,Harris2001,Wilde2005,Ruhe2006}%
, have been developed to observe the magnetization oscillations, i.e., the de
Haas-van Alphen (dHvA) effect in high-mobility 2DEG. Clear sawtooth dHvA
oscillations in the magnetization have been observed for LLs of filling
factors up to 52 \cite{Zhu2003}. More recently, a novel method has been used
by Prus et al. \cite{Prus2003} and Shashkin et al. \cite{Sha2006} to measure
the spin magnetization of 2DEG in silicon metal-oxide-semiconductor
field-effect transistors (MOSFETs). This method entails modulating the
magnetic field with an auxiliary coil and measuring the imaginary
(out-of-phase) component of the ac current induced between the gate and the
2DEG system, which is proportional to $\partial\mu/\partial B$ (where $\mu$ is
the chemical potential and $B$ is the magnetic field). Using the Maxwell
relation, $\partial\mu/\partial B$=$-\partial\mathcal{M}/\partial\mathcal{N}$,
one can then obtain the magnetization $\mathcal{M}$ by integrating the induced
current over the electron density, $\mathcal{N}$. Pauli spin susceptibility
has been observed to behave critically near the 2D metal-insulator transition,
in agreement with previous transport measurements \cite{Kra2004,Sha2005}. With
the similar method, Anissimova et al. \cite{Ani2006} have studied the
thermodynamic magnetization of a low-disordered, strongly correlated 2DEG in
silicon MOSFETs in perpendicular and tilted magnetic fields. By measuring
$\partial\mu/\partial B$ at noninteger filling factors, they have directly
determined the spectrum characteristics without any fitting procedures or parameters.

On the theoretical side, extensive studies of dHvA oscillations in the
magnetization of 2DEG have also been carried out
\cite{Sho1984,oppen,Bremme1999,Shar2004,Aldea2003,tan,imry}. In particular,
Bremme et al. \cite{Bremme1999} have investigated the influence of the edge
current on the dHvA oscillations in the magnetization of a 2DEG using a
\textit{spinless} single-particle approach. Sharapov et al. \cite{Shar2004}
have extensively discussed the dHvA oscillations of the magnetization in
planar system with the Dirac-like spectrum of quasiparticle excitation. In
addition, the magnetization oscillations as a function of the magnetic field
has also been theoretically studied in quantum dot systems
\cite{Aldea2003,tan,imry}. However, to our knowledge there are no detailed
treatments of the influence of edge states and the SOI on the magnetization in
2D systems.
In this paper, we study systematically the thermodynamic magnetization of a
2DEG system with edge states and SOI. In particular, we address the effects of
SOI and edge states on the Landau level (LL) structure, the chemical
potential, and the magnetization and its susceptibility to strong magnetic
field. Quantum oscillations in the magnetization of a 2DEG are well known to
be characterized by strictly $(1/B)$-periodic sawtoothlike oscillations with
an amplitude of $1$ effective Bohr magneton $\mu_{B}$ (=$e\hbar/2m^{\ast}$
with $m^{\ast}$ the effective electron mass) per electron. We will show that
the picture changes for the case in the presence of SOI and edge states. The
degeneracy of LL's plays an important role in the formation of dHvA
oscillations. The edge states lift partially the degeneracies of Landau levels
and lead to the change of both the center and the amplitude of the
sawtoothlike magnetization oscillation. The SOI mixes the spin-up and
spin-down states of neighboring LL's into two unequally spaced energy
branches. The inclusion of SOI changes the well-defined sawtooth behavior of
the dHvA oscillations in the magnetization. These results may be found useful
in the characterization of magnetic oscillations in two dimensional systems.

In Sec. II we review the exactly solvable cases of \textit{bulk} 2DEG with or
without Rashba SOI, and numerically solvable cases in the presence of both
edge states and Rashba SOI. In Sec. III we present the results for the
magnetization and the effects of SOI and/or edge states. The results on the
magnetic susceptibility are presented in section IV. Some concluding remarks
are given in Sec. V.


\section{Energy spectrum for 2DEG}

We consider a 2DEG with the Rashba coupling in the $x$-$y$ plane of an area
$L_{x}\times L_{y}$ subject to a perpendicular magnetic filed $\mathbf{B}%
$=$B\hat{z}$. The electrons are confined between $0$ and $L_{y}$ in the $y$
direction by an infinite potential wall, and its wave function is periodic
along the $x$ direction. We choose the Landau gauge $\mathbf{A}$%
=$-By\mathbf{\hat{x}}$. The Hamiltonian for a single electron of spin-$1/2$
with a Rashba coupling is given by
\begin{equation}
H_{0}=\frac{\vec{\Pi}^{2}}{2m^{\ast}}+\frac{\lambda}{\hbar}(\Pi_{x}\sigma
_{y}-\Pi_{y}\sigma_{x})-\frac{1}{2}g_{s}\mu_{B}B\sigma_{z}+V(y), \label{E1}%
\end{equation}
where $m^{\ast}$, $(-e)$, and $g_{s}$ are the electron's effective mass,
charge and effective magnetic factor respectively, $\mu_{B}$ is the Bohr
magneton, $\vec{\Pi}$=$\vec{p}+e\vec{A}/c$ is the kinetic operator, $\lambda$
is the Rashba coupling, and $\sigma_{\alpha}$ are the Pauli matrices.\ The
last term $V(y)$ is the lateral confining potential: $V(y)$=$0$ for
$0\leqslant y\leqslant L_{y}$ and infinite otherwise.\ Relevant quantities
related to the magnetic field are the cyclotron frequency $\omega_{c}%
$=$eB/m^{\ast}$ and the magnetic length $l_{b}$=$\sqrt{\hbar/eB}$.
The Rashba SOI in Eq. (1) stems from the structural inversion asymmetry (SIA)
introduced by a heterojunction or by surface or external fields. In
semiconductors with narrower energy gap (InGaAs, AlGaAs), this effect is
expected to be stronger \cite{Ras1960}. It has been shown experimentally that
the Rashba SOI can be modified up to $50\%$ by external gate voltages
\cite{Mil2003,Nit1997}.

Without considering the edge-state effect, the magnetization of a 2D
\textit{spinless} electron moving in a high perpendicular magnetic field is a
3D simplification that has been solved by Landau in a pioneering paper
published in 1930 \cite{Landau1930}. In that (2D) case, each state is
described by two quantum numbers $k$ and $n$. The quantum number $k$ denotes
the $x$ component of the electron momentum (scaled by $\hbar$) and is a
constant of motion. The other quantum number $n$ being the Landau-level index
describes different modes of a displaced linear oscillator of frequency
$\omega_{c}$. The LL's are%
\begin{equation}
E_{n}=(n+\frac{1}{2})\hbar\omega_{c} \label{E1A}%
\end{equation}
and the corresponding eigenstates are%
\begin{equation}
|n,k\rangle=\frac{1}{\sqrt{L_{x}}}e^{ikx}\frac{e^{-(y-y_{0})^{2}/2l_{b}^{2}}%
}{(\sqrt{\pi}2^{n}n!l_{b})^{1/2}}H_{n}(\frac{y-y_{0}}{l_{b}}), \label{E1B}%
\end{equation}
where $H_{n}(x)$ is the Hermite polynomial, and $y_{0}=l_{b}^{2}k$ is the
center of the cyclotron orbit. The degeneracy $N_{L}$ of each Landau level per
spin is given by $N_{L}$=$2\pi S/l_{b}^{2}$ with $S$=$L_{x}L_{y}$ being the
area of the sample. When the electron spin degrees of freedom is included but
the edge-state effect and spin-orbit coupling are excluded, the eigenstate is
characterized by three quantum numbers,%
\begin{equation}
|n,k,s\rangle=|n,k\rangle\chi_{s}, \label{E1C}%
\end{equation}
where $s=\pm1/2$ and $\chi_{\pm1/2}$ is the eigenstate of spin operator
$\hat{s}_{z}$ with eigenvalues $\pm\hbar/2$. In this case, the spin-split LL's
are
\begin{equation}
E_{n,s}=E_{n}-sg_{s}\mu_{B}B. \label{E1D}%
\end{equation}
Thus each LL is split into spin-up and spin-down branches.
When the edge channels are furthermore included but the SOI is excluded
\cite{Hal1982,Mac1984}, the eigenstate is still a product state of spin and
orbital degrees of freedom. However, the orbital part of the wave function is
no longer a form of Hermite function due to the confinement along $y$
direction. In this case, the eigenstate is given by $|n,k,s\rangle$=$\frac
{1}{\sqrt{L_{x}}}e^{ikx}\varphi_{n,y_{0}}(y)\chi_{s}$, where $\varphi
_{n,y_{0}}(y)$ obeys
\begin{equation}
\left[  -\frac{p_{y}^{2}}{2m^{\ast}}+\frac{1}{2}m^{\ast}\omega_{c}%
(y-y_{0})^{2}+V(y)\right]  \varphi_{n,y_{0}}(y)=\epsilon_{n}(y_{0}%
)\varphi_{n,y_{0}}(y). \label{E1E}%
\end{equation}
This equation has been solved by MacDonald et al. through properly applying
the boundary condition \cite{Mac1984}. The resulting eigenvalue spectrum of
Eq. (\ref{E1E}) has a form $\epsilon_{n}(y_{0})$=$[\nu_{n}(y_{0})+\frac{1}%
{2}]\hbar\omega_{c}$, where $\nu_{n}(y_{0})$ is numerically obtained by
requiring the wave function $\varphi_{n,y_{0}}(y)$ to vanish at the boundary
$y$=$0$, $L_{y}$. In this case, the spin-split LL's are given by%
\begin{equation}
E_{n,s}(y_{0})=[\nu_{n}(y_{0})+\frac{1}{2}]\hbar\omega_{c}-sg_{s}\mu_{B}B,
\label{E1F}%
\end{equation}
which as an example is illustrated in Fig. 1(a) as a function of the guiding
center $y_{0}$.

On the other side, when the Rashba SOI is included while the edge-state effect
is excluded, the spin-orbit coupling mixes the two spin components.
In this case, the energies of the two branches of states (denoted by $\pm$)
are given by
\begin{equation}
E_{n}^{\pm}=\hbar\omega_{c}\left(  n\pm\sqrt{(1-g)^{2}+8n\eta^{2}}\right)
\label{E1G}%
\end{equation}
for $n\geq1$, where we have defined $g$=$g_{s}m^{\ast}/2m_{e}$ and the
effective (dimensionless) Rashba coupling $\eta$=$\lambda m^{\ast}%
l_{b}/\hslash^{2}$. For $n=0$, there is only one single state with the energy
$E_{0}^{+}$=$\hbar\omega_{c}(1-g)$, which is the same as the lowest Landau
level without SOI. The corresponding eigenstates are given by
\begin{equation}
|n,y_{0},\pm\rangle=\left(
\begin{array}
[c]{c}%
\cos{\theta_{n}^{\pm}}|n,y_{0}\rangle\\
i\sin{\theta_{n}^{\pm}}|n-1,y_{0}\rangle
\end{array}
\right)  \label{E1H}%
\end{equation}
for $n\geq1$, where the parameters ${\theta_{n}^{\pm}}$ are given by
$\tan{\theta_{n}^{\pm}}$=$-u_{n}\pm\sqrt{1+u_{n}^{2}}$ with $u_{n}%
$=$(1-g)/\sqrt{8n}\eta$. For $n$=$0$, the single state is a product of the
ground-state oscillator mode $|0,k\rangle$ and eigenstate $\chi_{1/2}$ of
$\hat{s}_{z}$. Thus it is interesting to see that the ground state ($n$=$0$)
has the fully-polarized spin along the $z$ direction. In the excited states
the spin is tilted with an expectation value of its $z$ component
$\langle\sigma_{z}\rangle$=$\cos^{2}\theta_{n}^{\pm}-\sin^{2}\theta_{n}^{\pm}$
that decreases as $\lambda$ and $n$ increase.
A prominent feature is that the two branches of Landau levels $E_{n}^{+}$ and
$E_{n+1}^{-}$ cross each other at the values of $\eta$ satisfying
\begin{equation}
\sqrt{(1-g)^{2}+8n\eta^{2}}+\sqrt{(1-g)^{2}+8(n+1)\eta^{2}}=2. \label{E1I}%
\end{equation}
This degenerate behavior in the energy spectrum has been used to produce the
resonant spin-Hall current \cite{Shen2004,Wang2007}.

When both the edge states and the Rashba SOI are included, two interplayed
mixing mechanisms occur. One is from the coupling between different confining
orbital modes along the $y$ direction; the other is the mixing of the
eigenstates $\chi_{1/2}$ and $\chi_{-1/2}$ of the spin operator $s_{z}$. As a
result, the wave function for the final Hamiltonian (1) can be written in a
general form \cite{Reynoso}:%
\begin{equation}
\Psi_{n}(x,y)=\frac{1}{\sqrt{L_{x}}}e^{ikx}\varphi_{n}(y).\label{E1J}%
\end{equation}
Here $\varphi_{n}(y)$ are expanded in the basis of the infinite potential
well,%
\begin{equation}
\varphi_{n}(y)=\sqrt{\frac{2}{L_{y}}}\sum_{m}\sin\left(  \frac{\pi m}{L_{y}%
}y\right)  \left(
\begin{array}
[c]{c}%
a_{mn}\\
b_{mn}%
\end{array}
\right)  ,\label{E1K}%
\end{equation}
with $n$ being the Landau-level index and $m$ an integer. The Schr\"{o}dinger
equation $H_{0}\Psi_{n}$=$E_{n}(y_{0})\Psi_{n}$ leads to the following
equations for the spinors:
\begin{align}
&  \left[  A_{l\pm}-E_{n}\right]  \left(
\begin{array}
[c]{c}%
a_{ln}\\
b_{ln}%
\end{array}
\right)  \label{E1L}\\
&  =\sum_{m}\left[  i(F_{lm}+G_{lm})\sigma^{-}+i(F_{lm}-G_{lm})\sigma
^{+}-M_{lm}\right]  \left(
\begin{array}
[c]{c}%
a_{mn}\\
b_{mn}%
\end{array}
\right)  ,\nonumber
\end{align}
where $\sigma^{\pm}=(\sigma_{x}\pm i\sigma_{y})/2$, $\epsilon_{n}=E_{n}%
/\hbar\omega_{c}$.
The other parameters in Eq. (13) are defined as
\begin{align}
A_{l\pm} &  =\left(  \frac{\pi l}{2}\right)  ^{2}\left(  \frac{l_{b}}{L_{y}%
}\right)  ^{2}\mp\frac{g}{2},\label{E1M}\\
M_{lm} &  =\frac{1}{\pi^{3}}\left(  \frac{L_{y}}{l_{b}}\right)  ^{2}\int
_{0}^{\pi}d\theta\sin(l\theta)(\theta-\theta_{0})^{2}\sin(m\theta),\nonumber\\
F_{lm} &  =\frac{2\eta}{\pi^{2}}\frac{L_{y}}{l_{b}}\int_{0}^{\pi}d\theta
\sin(l\theta)(\theta-\theta_{0})\sin(m\theta),\nonumber\\
G_{lm} &  =2\eta\frac{l_{b}}{L_{y}}\int_{0}^{\pi}d\theta\sin(l\theta
)\frac{\partial}{\partial\theta}\sin(m\theta),\nonumber
\end{align}
where $\theta_{0}$=$y_{0}\pi/L_{y}$. We solve these equations in a truncated
Hilbert space disregarding the states with energies higher than the cutoff
energy. Typically we take a matrix Hamiltonian of dimension of a few hundred.
We increase the size of the Hilbert space by a factor 2 and find no change in
the results presented below. In all cases the width of the sample $L_{y}$ is
taken large enough to have the cyclotron radius $r_{c}$ smaller than $L_{y}%
/2$. The right and left edge states are then well separated in real space. A
typical energy spectrum is shown in Fig. 1(b). For $y_{0}\simeq L_{y}/2$ the
states are equal to the bulk states, except for exponential corrections. The
wave functions and the energy spectrum reproduces the above-discussed
\textit{bulk} results without edge states.
As $y_{0}$ approaches the sample edge, the effect of the confining potential
becomes important and it generates the $k$-dependent dispersion of the energy
levels \cite{Hal1982}, which has profound effects on magnetotransport and
magnetization properties. \begin{figure}[ptb]
\begin{center}
\includegraphics[width=0.8\linewidth]{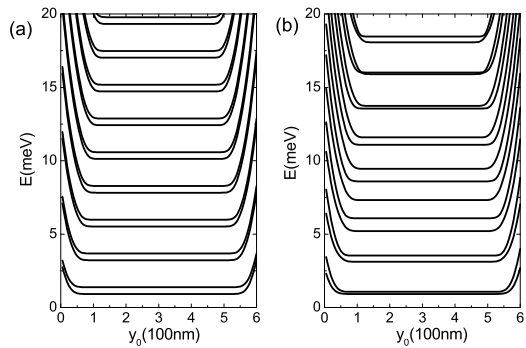}
\end{center}
\caption{The energy spectrum in units of meV versus the guiding center $y_{0}$
without (a) and with the Rashba spin-orbit couplings (b), respectively. In
both figures, the parameters are chosen as $m^{\ast}=0.05m_{e}$, $L=600$ nm,
$B=1$ T, and $g_{s}=4$. The Rashba SOI strength in (b) is set as $\lambda
=15$meV nm.}%
\end{figure}

\section{The Magnetization of 2DEG}

We begin with a briefly review of the standard derivation of the 2DEG
magnetization from the free energy.
The magnetization density is the derivative of the Helmholtz free energy
density with respect to $B$ at fixed electron density $\mathcal{N}$ and
temperature $T$, $M$=$-(\partial F/\partial B)|_{\mathcal{N},T}$. For the
present model with the LL's $E_{n,s}(y_{0})$, the Helmholtz free energy
density is given by%
\begin{align}
F(B,T) &  =\mu\mathcal{N}-\frac{1}{L_{y}}\frac{N_{\nu}}{\beta}\int_{0}^{L_{y}%
}dy_{0}\sum_{n,s}\ln\left\{  1+e^{\beta\lbrack\mu-E_{n,s}(y_{0})]}\right\}
\label{E2}\\
&  \equiv\mu\mathcal{N-}\frac{1}{\beta}\int dED(E,B)\ln\{1+e^{\beta(\mu
-E)}\},\nonumber
\end{align}
where $\beta$=$1/k_{B}T$, $N_{\nu}$=$N_{L}/S$, and $\mu$ is the chemical
potential. Note that we have defined in the above equation the density of
states (DOS) per area%
\begin{align}
D(E,B) &  =\frac{1}{L_{x}L_{y}}\sum_{n,k,s}\delta(E-E_{n,s}(y_{0}%
))\label{E2A}\\
&  =\frac{N_{\nu}}{L_{y}}\sum_{n,s}\int_{0}^{L_{y}}dy_{0}\delta(E-E_{n,s}%
(y_{0})),\nonumber
\end{align}
where we have replaced $k$-sum with $\frac{L_{x}}{2\pi}\int dk$ and used the
relation $y_{0}$=$l_{b}^{2}k$. The explicit inclusion of the DOS in the
expression can be utilized to take into account the impurity effect, which
broadens the LL's into Gaussian or Lorentzian in shape. For simplicity we did
not consider the broadening effect in this paper. In the absence of edge
states, the LL's $E_{n,s}(y_{0})$ are uniform in space and thus Eq. (\ref{E2})
reduces to
\begin{equation}
F(B,T)=\mu\mathcal{N}-\frac{N_{\nu}}{\beta}\sum_{n,s}\ln\left\{
1+e^{\beta(\mu-E_{n,s})}\right\}  .\label{E2B}%
\end{equation}
The $B$ dependent chemical potential $\mu$ is connected to the experimentally
accessible electron density $\mathcal{N}$ via the local density of states
(DOS). In the clean sample limit this is written as
\begin{equation}
\mathcal{N}=\frac{N_{\nu}}{L_{y}}\int_{0}^{L_{y}}dy_{0}\sum_{n,s}f_{ns}\left(
y_{0}\right)  ,\label{E2C}%
\end{equation}
where $f_{ns}\left(  y_{0}\right)  $=$\frac{1}{e^{\beta\left[  E_{n,s}%
(y_{0})-\mu\right]  }+1}$ is the Fermi distribution for the spin-split LL's
$E_{n,s}(y_{0})$. From Eq. (\ref{E2B}) the magnetization density becomes%
\begin{align}
M &  =\sum_{n,s}\left\{  -N_{\nu}\int_{0}^{L_{y}}\frac{dy_{0}}{L_{y}}%
f_{ns}\left(  y_{0}\right)  \frac{\partial E_{n,s}(y_{0})}{\partial B}\right.
\nonumber\\
&  \left.  +\frac{e}{h}\frac{1}{\beta}\int_{0}^{L_{y}}\frac{dy_{0}}{L_{y}}%
\ln\left\{  1+e^{\beta\left[  \mu-E_{n,s}(y_{0})\right]  }\right\}  \right\}
\nonumber\\
&  \equiv M^{(0)}+M^{(1)}.\label{E2D}%
\end{align}
One can see that the magnetization consists of two parts. The first
part $M^{(0)}$ is the conventional contribution from the $B$
dependence of the LL's and thus denotes a diamagnetic response. The
second part $M^{(1)}$ comes from the $B$ dependence of the level
degeneracy factor $N_{\nu}$, thus describing the effect of the
variation of the density of states upon the magnetic field and
denoting a paramagnetic contribution to the total magnetization.
Obviously, $M^{(0)}$ is negative while $M^{(1)}$ is positive, the
net result is an oscillation of the total magnetization $M$ between
the negative and positive values as a function of $B$. At zero
temperature, the expression for $M$ reduces to a sum over all
occupied LL's:
\begin{align}
M  & =\sum_{n,s}^{\text{occ}}\left\{  -N_{\nu}\int_{0}^{L_{y}}\frac{dy_{0}%
}{L_{y}}\frac{\partial E_{n,s}(y_{0})}{\partial B}\right.  \label{E2E}\\
& \left.  +\frac{e}{h}\int_{0}^{L_{y}}\frac{dy_{0}}{L_{y}}[\mu_{0}%
-E_{n,s}(y_{0})]\right\}  ,\nonumber
\end{align}
where the sum runs over all occupied states and $\mu_{0}$ is the
zero-temperature chemical potential (Fermi energy).\begin{figure}[ptb]
\begin{center}
\includegraphics[width=0.6\linewidth]{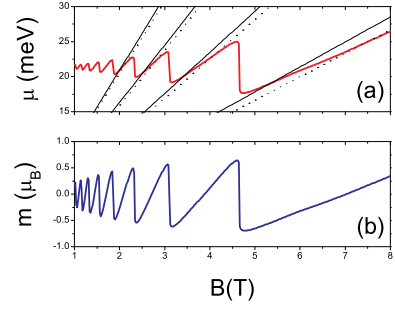}
\end{center}
\caption{(Color online) Magnetic field dependence of (a) Chemical potential
$\mu$ (thick line) and Landau levels (dashed line: spin down; solid line: spin
up) and (b) magnetization $m$ (per electron) for the 2DEG without edge states
and spin-orbit coupling. The other system parameters are $\mathcal{N}%
=4.5\times10^{-3}/$nm$^{2}$, $g_{s}=4$, $m=0.05m_{e}$, for the inversion
heterostructure In$_{0.53}$Ga$_{0.47}$As/In$_{0.52}$Al$_{0.48}$As, and $T$=3.5
K.}%
\end{figure}\begin{figure}[ptbptb]
\begin{center}
\includegraphics[width=0.6\linewidth]{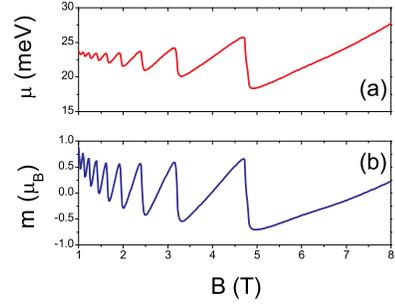}
\end{center}
\caption{(Color online) The chemical potential $\mu$ (a) and the magnetization
per electron $m$ (b) in units of $\mu_{B}$ in the sample with edges as a
function of the external magnetic field $B$. The Rashba coupling $\lambda=0$.
The sample size $L=600$nm. The other parameters are the same as those in Fig.
2.}%
\end{figure}

For comparison, let us start from the conventional result for the
\textit{bulk} 2DEG without SOI and edge-state effects. In this case, both the
chemical potential $\mu$ [Fig. 2(a)] and the magnetization (per electron) $m$
[Fig. 2(b)] display the well-known sawtooth behavior with varying the magnetic
field. At zero temperature, the explanation of the dHvA oscillation can be
given with the help of the filling factor $\nu$=$\mathcal{N}/N_{\nu}$, which
measures the number of occupied LL's and is an integer when all the available
states in the $\nu$ lowest Landau levels are filled. At these integer values
the 2DEG is incompressible and the chemical potential jumps discontinuously
between two adjacent LL's by an amount of $\Delta\mu_{0}$=$\hbar\omega_{c}%
$=$2\mu_{B}B$ ($\mu_{B}$=$eB/2m^{\ast}$), which defines the incompressibility
gap. Note that the abrupt jump in the dHvA oscillation is on the high magnetic
field side of the sawtooth, which is special for our present choice of the
thermodynamic system. If the system is constrained to have constant chemical
potential, then the jump in the dHvA oscillation will move to the low magnetic
field side of the sawtooth, which has been confirmed by Meinel \textit{et al}.
\cite{Mei1999} in an experiment with the electron density $\mathcal{N}%
$\ modulated by applying a gate voltage to the sample. The zero-temperature
behavior of the magnetization curve can be seen by Eq. (\ref{E2E}), which in
the absence of edge states is%
\begin{equation}
M=\frac{e}{h}\sum_{n,s}^{\text{occu.}}(\mu_{0}-2E_{n,s}). \label{E2F}%
\end{equation}
From above equation, one can easily derive a simplified Maxwell relation
$\frac{\Delta M}{\mathcal{N}}$=$\frac{\Delta\mu}{B}$. Thus the discontinuous
jump $\Delta M$ is related to the discontinuity in the chemical potential
$\Delta\mu$, i.e., the Landau energy gap at even and the Zeeman gap at odd
filling factors, according to $\Delta\mu=\Delta M\cdot B$. When the magnetic
field increases within an odd filling factor $\nu$, the Fermi energy $\mu_{0}$
and LL's also increase with $B$. As a result, the magnetization also increases
rapidly as a function of $B$, then evolves a maximum at adjacent filling
factor $\nu-1$ and suddenly jumps to a negative value. The zero-temperature
magnetization jump at these even filling factors are given by the above
Maxwell relation, $\Delta M$=$2\mu_{B}$, independent of the magnetic field and
spin-splitting. At finite temperature as shown in Fig. 2 ($T$=$3.5$ K), the
oscillation amplitude of the magnetization increases with increasing the
magnetic field. This fact is due to that the influence of finite temperature
in this case (i.e., no SOI and no edge states in the 2DEG) is merely to reduce
the oscillation amplitude and the discontinuities in $\mu$ and $M$ via the
smearing of the Fermi-Dirac distribution. Another fact revealed in Fig. 2 is
that the inclusion of the Zeeman splitting in the LL's does not change the
dHvA oscillation modes of the physical quantities. It is due to this fact that
the LL's are usually treated to be spin degenerate in previous work.

\begin{figure}[ptb]
\begin{center}
\includegraphics[width=0.8\linewidth]{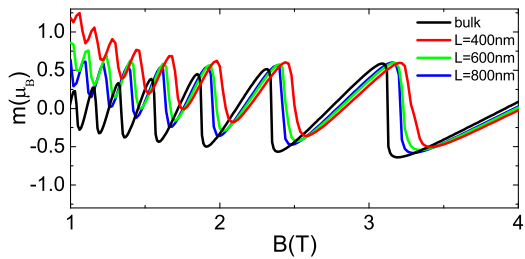}
\end{center}
\caption{(Color online) The magnetization per electron $m$ in units of
$\mu_{B}$ in the sample with edges as a function of the external magnetic
field $B$. The Rashba coupling $\lambda=0$. The sample size $L=400,600,800$nm.
The other parameters are the same as those in Fig. 2.}%
\end{figure}

\begin{figure}[ptb]
\begin{center}
\includegraphics[width=1.0\linewidth]{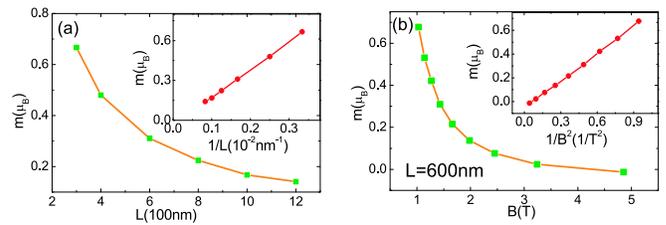}
\end{center}
\caption{(Color online) (a) The dependence of the center of dHvA oscillation
on the size of the sample. The magntic field $B$ is chosen aroud $1.2$T. (b)
The $B$ dependence of the center of dHvA oscillations. In both figures the
Rashba SOI is neglected.}%
\end{figure}

\begin{figure}[ptb]
\begin{center}
\includegraphics[width=0.8\linewidth]{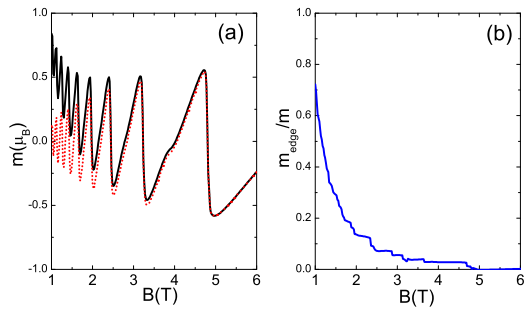}
\end{center}
\caption{(Color online) (a) Red curve: The bulk contribution of the
magnetization $m$ per electron (in units of $\mu_{B}$); Black curve: The total
magnetization per electron $m$ (in units of $\mu_{B}$). The Rashba coupling
$\lambda=0$. (b) The ratio of edge states contribution of magnetization to the
total magnetization. $L=600$nm.}%
\end{figure}

Now let us see the edge-state effects on the magnetization. Fig. 3 shows the
influence of the edge states on the oscillations of chemical potential and
magnetization (dHvA oscillations) with magnetic field. The most prominent
feature brought by the edge states is that the center of the dHvA oscillations
is now dependent on the magnetic field. In particular, for the field less than
$1$ Tesla, the oscillatory magnetization is always positive in sign. Another
feature shown in Fig. 4 is that the oscillation amplitude decreases with
decreasing the sample size. As is known, the origin of dHvA oscillations is
the degeneracy of Landau levels. The edge states with dispersion lead to edge
current which not only is crucial for the quantum Hall effects, but also very
important for the magnetization \cite{Bremme1999}. The dispersion of the edge
states partially lift the degeneracy of the Landau levels. Thus the edge
states tend to destroy the dHvA oscillations. Therefore it leads to the
decreasing of oscillation amplitude as shown in Fig. 4. The upshift of the
center of dHvA oscillations may be understood in the following way. For the
effects from the edge states, what really matters is the ratio of two
important length scales: the magnetic length $l_{b}$ and the system size
$L_{y}$. The decreasing of $L_{y}$ is equivalent to the increasing of $l_{b}$,
i.e., decreasing of $B$ or $\omega_{c}$. As seen in Eq. (6), in the case with
smaller $L_{y}$ or weaker $B$, the second term becomes less important and the
eigenenergy is less sensitive to the magnetic field for states with $y_{0}$
near the edges. From Eq. (\ref{E2E}), one sees that the second term overcomes
the first term and leads to the upshift of the center of dHvA oscillations.
The smaller the system size is, the more profound effects the edge states lead
to, as seen in Fig. 4 for both the center and amplitude of the dHvA
oscillations. Fig. 5(a) shows quantitatively the system size dependence of the
shift of the oscillation center. It has the dependence $1/L_{y}$. Roughly, the
contribution of the edge states is proportional to the number of edge states
(as also seen from Eq.(\ref{E2E})), which is proportional to $\nu r_{c}/L_{y}%
$, where the cyclotron radius $r_{c}=\sqrt{\nu}l_{b}$ \cite{Reynoso}, with the
number of the occupied Landau levels $\nu\sim1/l_{b}^{2}$. Thus the center of
dHvA oscillations is proportional to $l_{b}^{4}/L_{y}=1/B^{2}L_{y}$. The $B$
and $L_{y}$ dependence is clearly seen in Figs. 4 and 5. To see more
explicitly the contribution from edge states and bulk states, we plot the
total magnetization and the contribution from bulk states in Fig. 6(a).
The contribution from the edge states is obtained from Eq. (\ref{E2D}) by
summing over terms from edge states, with $|y_{0}|<r_{c}$ or $|L-y_{0}|<r_{c}%
$. The rest contribution is from bulk states. There is no upshift of the
magnetization oscillation center for the part from bulk states. It shows
explicitly that the upshift of the center of dHvA oscillations is due to the
existence of edge states. Fig. 6(b) shows the dependence of edge states
contribution on the magnetic field. The contribution from edge states
increases as decreasing the magnetic field, or equivalently decreasing the
sample size as one expects.



When the Rashba SOI is introduced, the filling factor $\nu$ is not linearly
proportional to the inverse of the external field $B$, and there is a energy
competition between the Zeeman coupling and the Rashba coupling. Also, due to
the entanglement between the orbital and spin degrees of freedom, it is
difficult to distinguish their separate contributions to the total
magnetization. These factors make the physical picture of the dHvA
oscillations to change fundamentally, as shown in Fig. 7(a) for chemical
potential $\mu$ and Fig. 7(b) for magnetization $M$ as functions of $B$. One
can see from these two figures that the Rashba SOI has no visible influence on
the magnetic oscillations of the quantities $\mu$ and $M$ at large values of
$B$, where the Zeeman and spin-orbit coupling splitting are small compared to
the Landau level splitting. At low magnetic field, however, the SOI modulation
of the magnetic oscillations becomes obvious, which can be clearly seen by the
enlarged plots of $\mu$ and $M$ in Fig. 7(c) and 7(d) respectively for $B$
less than $2.4$ T. For comparison, we also re-plot in Figs. 7(c)-(d) the cases
without Rashba SOI. One can see from these two figures that the SOI brings
about two new features at low magnetic field: (i) The sawtooth structure is
inversed, i.e., the location of peaks in $\mu$ and $M$ with SOI correspond to
the valleys without SOI. This inversion is due to the different LL's in the
two cases. (ii) The oscillation mode is prominently modulated by SOI and a
beating pattern appears. This beating behavior in the oscillations are due to
the fact that the LL's $E_{n}^{+}$ and $E_{n}^{-}$ are now unequally spaced
due to the presence of SOI.\begin{figure}[ptb]
\begin{center}
\includegraphics[width=1.0\linewidth]{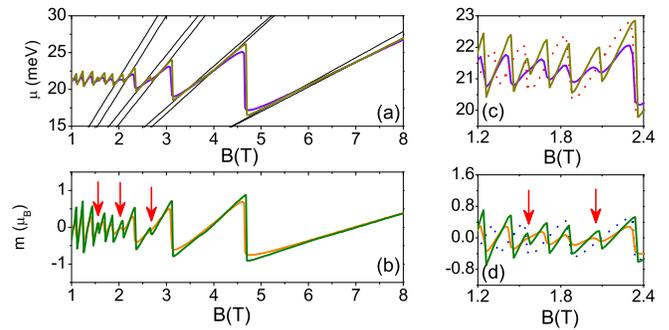}
\end{center}
\caption{(Color online) (a) The Chemical potential $\mu$ and (b) the
magnetization $m$ (per electron) for the 2DEG under different temperatures
$T=3.5$K and $1$K with Rasha spin-orbit interaction in the absence of edge
states. the Rashba coupling is taken to be $\lambda$=$15$ meV nm. The other
parameters are same as those in Fig. 2. The detailed oscillations of $\mu$ and
$m$ at low magnetic fields are illustrated in (c) and (d) respectively (solid
lines), in comparison with the case of $\lambda$=$0$ (dotted lines). }%
\end{figure}\begin{figure}[ptbptb]
\begin{center}
\includegraphics[width=1.0\linewidth]{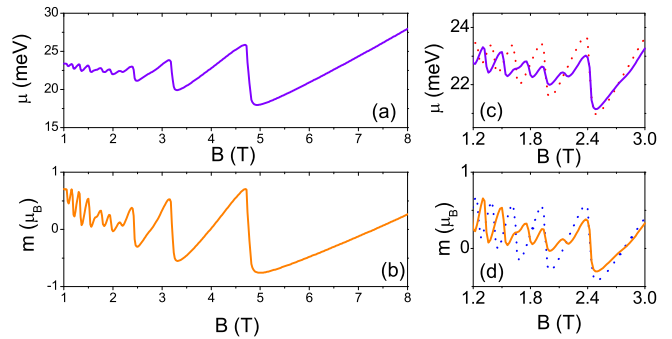}
\end{center}
\caption{(Color online) The chemical potential $\mu$ (a,c) and the
magnetization $m$ (per electron in units of $\mu_{B}$) (b,d) in the sample
with edges as a function of the external magnetic field $B$. The Rashba
coupling $\lambda=0$ (a,b) and $\lambda=15$meV nm (c,d). In four figures, the
sample size $L=600$nm. The temperature $T=3.5$ K.}%
\end{figure}

Another effect caused by the Rashba spin-orbit coupling is that there are weak
peaks appearing in the chemical potential and the magnetization oscillation at
low magnetic field, as shown with the arrows in Fig. 7.
In the 2DEG sample without edges, the weak peaks will appear in the
magnetization once the temperature is sufficient low, for example, $T$=$1$K.
These weak peaks have been observed recently by Schaapman et
al.\cite{Schaapman2003} when they measured the magnetization of a dual-subband
2DEG, confined in a GaAs/AlGaAs heterojection, and by Zhu et al.\cite{Zhu2003}
when they measured the magnetization of high-mobility 2DEG. These peaks is so
weak that they will disappear when the temperature is much higher.

Now let us see the SOI effect superposed on the edge-state effect. Figures
8(a)-(b) show the chemical potential $\mu$ and magnetization $M$ respectively
in the presence of both SOI\ and edge states. One can see from these two
figures that at large values of $B$, where the Zeeman and spin-orbit coupling
splitting are small compared to the Landau level splitting, and thus the
Rashba SOI has no big influence on the magnetic oscillations of the quantities
$\mu$ and $M$. Neither do the edge states, since the cyclotron radius is much
smaller than the system size. At low magnetic field, however, both edge states
and Rashba SOI change the pattern of dHvA oscillations, as clearly seen in
Figs. 8(a), 8(b), and the enlarged plots of $\mu$ and $M$ in Figs. 8(c) and
(d) respectively for $B$ less than $2.4$ T. For comparison, we also re-plot in
Figs. 8(c)-(d) the cases without Rashba SOI. One can see that the total
effects are the superposition of effects from both the edge states and Rashba
SOI.

\section{Magnetic susceptibility of 2DEG}

Now we turn to study the magnetic susceptibility $\chi(B)$ for the 2DEG. From
Eq. (\ref{E2D}), one obtains the expression for $\chi(B)$ as follows:%
\begin{align}
\chi\left(  B\right)   &  =\frac{\partial M}{\partial B}\label{E3}\\
&  =\sum_{n,s}\left\{  \frac{e}{h}\int_{0}^{L_{y}}\frac{dy_{0}}{L_{y}}%
f_{ns}(y_{0})\left(  \frac{\partial\mu}{\partial B}-2\frac{\partial
E_{ns}(y_{0})}{\partial B}\right)  \right.  \nonumber\\
&  -N_{\nu}\int_{0}^{L_{y}}\frac{dy_{0}}{L_{y}}\left[  \frac{\beta}{4\cosh
^{2}\frac{\beta\left[  E_{n,s}(y_{0})-\mu\right]  }{2}}\left(  \frac{\partial
E_{ns}(y_{0})}{\partial B}\right)  ^{2}\right.  \nonumber\\
&  \left.  \left.  +f_{ns}(y_{0})\frac{\partial^{2}E_{ns}(y_{0})}{\partial
B^{2}}\right]  \right\}  ,\nonumber
\end{align}
which at zero temperature reduces to%
\begin{align}
\chi\left(  B\right)    & =\sum_{n,s}^{\text{occu.}}\left\{  \frac{e}{h}%
\int_{0}^{L_{y}}\frac{dy_{0}}{L_{y}}\left(  \frac{\partial\mu_{0}}{\partial
B}-2\frac{\partial E_{ns}(y_{0})}{\partial B}\right)  \right.  \label{E3A}\\
& \left.  -N_{\nu}\int_{0}^{L_{y}}\frac{dy_{0}}{L_{y}}\frac{\partial^{2}%
E_{ns}(y_{0})}{\partial B^{2}}\right\}  .\nonumber
\end{align}
For the ideal noninteracting 2DEG without the edge states, the second term in
Eq. (\ref{E3A}) is zero due to the fact that both the Landau and the Zeeman
splittings of the energy spectrum are linear in $B$. Thus in this case (zero
temperature and no edge states) the magnetic susceptibility is simply written
as
\begin{equation}
\chi\left(  B\right)  =\frac{e}{h}\sum_{n,s}^{\text{occu.}}\left(
\frac{\partial\mu_{0}}{\partial B}-2\frac{\partial E_{ns}}{\partial B}\right)
.\label{E3B}%
\end{equation}
Furthermore, if the SOI is disregarded, the derivative $\partial\mu
_{0}/\partial B$ in quantizing magnetic fields (except at even integer filling
factors) is equal to
\begin{equation}
\frac{\partial\mu_{0}}{\partial B}=\mu_{B}\left[  (n+\frac{1}{2})\frac{2m_{e}%
}{m^{\ast}}\mp\frac{1}{2}g_{s}\right]  .\label{E3C}%
\end{equation}
Then Eq. (\ref{E3B}) clearly shows Landau diamagnetic ($\chi<0$) at Landau
gaps and Pauli paramagnetic ($\chi>0$) responses at Zeeman gaps.

Let us first see the SOI effect on the magnetic susceptibility. Figure 9(a)
shows $\chi\left(  B\right)  $ as a function of $1/B$ for the edgeless 2DEG.
One can see from Fig. 9(a) that there are a series of equal-distance resonance
peaks appearing in the magnetic susceptibility $\chi$ with the magnetic field
with/without the Rashba coupling $\lambda$. With the magnetic filed
increasing, the magnitude of the resonance susceptibility increases. Similar
to that of the magnetization, the explanation of the resonance peaks of the
magnetic susceptibility without Rashha coupling also needs the help of the
filling factor $\nu\sim1/B$. When $\nu$ increases to an integer value, the
2DEG is incompressible. At this time, all the available states in the $\nu$
lowest Landau levels are filled. Upon increasing the inverse magnetic field
$1/B$, the electrons are transferred to the next Landau level. Thus the
chemical potential changes discontinuously (See Fig. 2(a)) and the
magnetization jumps (See Fig. 2(b)). Therefore there is a corresponding
resonance peak appearing in the magnetic susceptibility $\chi$. When the
Rashba spin-orbit coupling is introduced, it will destroy the simply linear
relation between the filling factor $\nu$ and the inverse magnetic field
$1/B$. When the value of $1/B$ is larger, the SOI effect is more evident (See
the red line in the Fig. 9(a)).\begin{figure}[ptb]
\begin{center}
\includegraphics[width=1.0\linewidth]{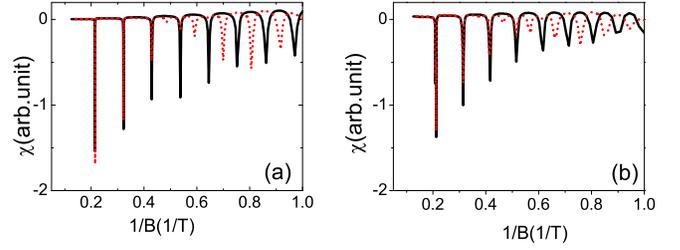}
\end{center}
\caption{(Color online) The magnetic susceptibility $\chi$ in the sample (a)
\emph{without} and (b) \emph{with} edges as a function of the inversed
magnetic field $1/B$. The black (real) line is corresponding to the case of
the Rashba SOI strength $\lambda=0$, while the red (dotted) line is
corresponding to the case of $\lambda=15$meV nm. In both figures, the electron
density and the temperature are respectively set as $n_{s}=4.5\times10^{-3}%
$/nm$^{2}$ and $T=3.5$K. The system size in (b) $L=600$nm.}%
\end{figure}

Figure 9(b) plots the magnetic susceptibility $\chi$ in a 2DEG sample
\emph{with} edges as a function of the inversed magnetic field $1/B$.
Comparing the black lines in Figs. 9(a) and 9(b), which corresponding to the
case without Rashba spin-orbit coupling, one can find that
the resonant (for magnetic susceptibility) magnetic field shifts to larger
values for the case with edges. We have obtained the conclusion that the edges
not only affect the center of the dHvA oscillation, but also the values of the
magnetic field when the magnetization has the discontinuous change. As a
consequence, the conditions of the susceptibility (defined as $\chi$=$dM/dB$)
having resonance are the same with those of the magnetization. When the Rashba
spin-orbit coupling is introduced, similar to the case in the sample without
edges, the Rashba coupling changes the resonance condition. When the value of
$1/B$ is much larger, the change is much larger as seen in Fig. 9(b).

\section{Summary}

In summary, we have systematically studied the dHvA oscillations of the
magnetization and its susceptibility for the 2DEG with the edges states and
SOI included in the system.

We find that the edge states and Rashba SOI play important roles when the
external magnetic field is small. The edge effect prominently changes the
oscillation center and oscillation amplitude. The dHvA oscillation will change
the sawtoothlike form if the Rashba coupling introduced, no matter the sample
is with or without edges. The total effects are the superposition of effects
from both edge states and SOI.

\begin{acknowledgments}
This work was supported by NSFC under Grants No. 10604010, No. 60776063, No.
10744004, and No. 10874020, and by the National Basic Research Program of
China (973 Program) under Grant No. 2009CB929103.
\end{acknowledgments}

\end{document}